%% file: jpsipA-alicepreprint_240114.tex
\documentclass[ALICE,manyauthors]{cernphprep}
\usepackage{hyperref}
\usepackage{cite}

\usepackage{rotating}

\begin{document}%
%
%
\begin{titlepage}
\PHnumber{2013-163}      
\PHdate{24 August 2013}            
%
%
\title{J/$\mathbf{\psi}$ production and nuclear effects in \mbox{p-Pb}
collisions at $\mathbf{\sqrt{s_{\rm NN}}}$=5.02 TeV}

\ShortTitle{J/$\mathbf{\psi}$ production and nuclear effects in \mbox{p-Pb} collisions at $\mathbf{\sqrt{s_{\rm NN}}}$ = 5.02 TeV}
%
\Collaboration{ALICE Collaboration%
         \thanks{See Appendix~\ref{app:collab} for the list of collaboration 
                      members}}
\ShortAuthor{ALICE Collaboration}      
\begin{abstract}
Inclusive J/$\psi$ production has been studied with the ALICE detector in \mbox{p-Pb} collisions at the nucleon--nucleon center of mass energy $\sqrt{s_{\rm NN}}$ = 5.02 TeV at the CERN LHC.
The measurement is performed 
in the center of mass rapidity domains $2.03<y_{\rm cms}<3.53$ and $-4.46<y_{\rm cms}<-2.96$, down to zero transverse momentum, studying the $\mu^+\mu^-$ decay mode. In this paper,  
the J/$\psi$ production cross section and the nuclear modification factor
$R_{\rm pPb}$ for the rapidities under study are presented. While at forward
rapidity, corresponding to the proton direction, a suppression of the J/$\psi$ yield with respect to binary-scaled pp 
collisions is observed, in the backward region no suppression is present. 
The ratio of the forward and backward yields is also measured differentially in rapidity and transverse momentum. 
Theoretical predictions based on nuclear shadowing, as well as on models 
including, in addition, a contribution from partonic energy loss, are in fair 
agreement with the experimental results.
\end{abstract}
\end{titlepage}
\setcounter{page}{2}
%
\input{jpsipA_240114_ap.tex}               
\newenvironment{acknowledgement}{\relax}{\relax}
\begin{acknowledgement}
\section*{Acknowledgements}
\input{acknowledgements_march2013.tex}    
\end{acknowledgement}
%

%
\newpage
\appendix
\section{The ALICE Collaboration}
\label{app:collab}
\input{Alice_Authorlist_2013-Aug-20_new-CERNPREP.tex}  
\end{document}

%% file: jpsipA_240114_ap.tex
The production of charmonia, bound states of $c$ and $\overline{c}$ quarks, is the object of intense theoretical and experimental 
investigations~\cite{Bra11}. As of today, their production mechanism in pp collisions is described by models based on Quantum Chromodynamics (QCD). 
In particular, in the NRQCD (non-relativistic QCD) approach~\cite{Bod95}, charmonium production is seen as a two-step process which includes 
the creation of the $c\overline{c}$ pair in a hard scattering, described perturbatively, and the subsequent evolution of the pair towards a 
bound state with specific quantum numbers, which is modeled in a non-perturbative way. In this model, the evolving $c\overline c$ pair can be in a color-singlet (CS) as well as in a color-octet (CO) state, with the strength of the CO amplitude contributions being controlled by non-perturbative factors, extracted by fits to experimental data (see~\cite{But11} for a recent implementation based on HERA, RHIC and LHC results). 

Several initial/final-state effects related to the presence of cold nuclear matter can
influence the observed charmonium yields in proton-nucleus collisions. Concerning the
initial state, the kinematical distributions of partons in nuclei are different from
those in free protons and neutrons (nuclear shadowing~\cite{Aub83,Esk09,deF12,deF04,Hir07}), affecting the production cross section of the $c\overline{c}$ pair. Therefore, charmonium production measurements help in constraining the nuclear parton distribution functions for gluons, which at hadron collider energies dominate the production process. Alternatively, when the production process is dominated by low-momentum gluons, i.e. carrying a small fraction $x_{\rm Bj}$ (Bjorken-$x$) of the momentum of the hadron, the Color-Glass Condensate (CGC) effective theory~\cite{Kha06,Fuj06} describes the nucleus as a dense (saturated) partonic system, and gives, once it is combined with a specific pp production model,  predictions for the charmonium yields. In addition, the initial parton inside the proton may suffer energy loss before the hard collision producing the $c\overline{c}$ pair takes place, shifting in this way the center-of-mass energy $\sqrt{s}$ of the partonic collision~\cite{Gav92, Bro93, Arl12}. This effect can result in a suppression of charmonia at large longitudinal momentum. 

Once created, the evolving $c\overline{c}$ pair needs a finite amount of time (up to several fm/$c$ in the nucleus rest frame) to
form the final-state charmonium. It may, therefore, interact with the nuclear matter and possibly break-up, with the break-up cross section being sensitive to the nature (color-octet or singlet) of the intermediate state~\cite{Vog02,Kop01,Vog12}. In addition, the final state may also experience energy loss, leading to a reduction of the pair momentum~\cite{Vog00}. It is also worth noting that recent approaches to the parton energy loss effect led to the hypothesis of a coherent energy loss which cannot be factorized into initial and final-state contributions~\cite{Arl12}.

Experimental studies have been carried out at various collision energies, for nuclei of different sizes, and differentially in rapidity ($y$) and transverse momentum ($p_{\rm T}$). These studies allow the amount of nuclear matter crossed by the $c\overline{c}$ pair to be varied, modifying the environment of its  
evolution, as well as the initial parton kinematics. In this way, further constraints to theoretical models can be provided.

Finally, the small size ($< 1$ fm) and large binding energy (several hundred MeV) of some of the charmonium states make them ideal probes of the strongly interacting matter created in ultrarelativistic heavy-ion collisions, which at
sufficiently high energy density may become a Quark-Gluon Plasma (QGP). A suppression of charmonium production was predicted as a signature of the phase 
transition to a QGP~\cite{Mat86} and observed at SPS~\cite{Ale05,Arn07,Arn09} ($\sqrt{s_{\rm NN}}\sim$ 20 GeV) and RHIC~\cite{Ada11,Abe09} ($\sqrt{s_{\rm NN}}=$ 200 GeV), and more recently at the 
LHC~\cite{Abe12,Abe13J,Cha12,Aad11} ($\sqrt{s_{\rm NN}}=$ 2.76 TeV). However, in
such collisions, suppression mechanisms related to initial-state effects and/or interaction of charmonia with cold nuclear matter have been verified to play a role~\cite{Ale03,Ada11b}. Results on proton-nucleus collisions are therefore  essential to calibrate and 
disentangle these effects in order to allow a quantitative determination of the QGP-related suppression in nucleus-nucleus collisions. 

A large amount of experimental results is available today for the production of J/$\psi$, the most strongly bound charmonium 
state decaying into dileptons, in proton-nucleus collisions. 
Fixed-target experiments at SPS~\cite{Ale06,Arn12}, Tevatron~\cite{Lei00} and HERA~\cite{Abt09}, as well as collider experiments at RHIC~\cite{Ada13} have 
investigated J/$\psi$ production in large kinematic ranges in the Feynman-$x$ ($x_{\rm F}=2p_{\rm L}/\sqrt{s}$, where $p_{\rm L}$ is the longitudinal momentum) and $p_{\rm T}$ variables. Among the main features of the 
results, a suppression of the J/$\psi$ yield, relative to the one in proton-proton collisions, has been observed, which increases at high $x_{\rm F}$ (corresponding to forward $y$). In addition, at fixed $x_{\rm F}$ the suppression decreases with increasing $\sqrt{s_{\rm NN}}$~\cite{Arn12}. Finally, the suppression is found to steadily decrease~\cite{Lei00,Abt09} as a function of $p_{\rm T}$. Several attempts have been made to describe these observations  theoretically, based on the different physics mechanisms described above~\cite{Vog00}. 
Although some features of the data are correctly reproduced, a quantitative understanding has not yet been reached. 

In this context, data from the LHC can bring new information and help to clarify the situation. On the production side, very small $x_{\rm Bj}$ values, down to $\sim10^{-5}$, can be accessed, allowing the gluon distributions to be studied in a previously unexplored kinematic range. On the other hand, the large Lorentz $\gamma$-factor of the $c\overline{c}$ pair, in particular at forward $y$, makes its crossing time through nuclear matter very short. In this kinematic range one may therefore expect a negligible $c\overline c$ break-up probability and the yield to be dominated by initial-state effects and possibly energy loss. In addition, proton-nucleus results
are essential, as it was the case at lower energies~\cite{Ale05}, in the interpretation of the J/$\psi$ suppression effects seen in Pb-Pb collisions at the LHC.

In this paper, we present the first results from ALICE on inclusive J/$\psi$ production at the LHC in \mbox{p-Pb} collisions at $\sqrt{s_{\rm NN}}=5.02$ TeV. These results have been obtained with  measurements in the $\mu^+\mu^-$ decay channel in the muon spectrometer, which covers the pseudorapidity range $-4<\eta_{\rm lab}< -2.5$. The muon spectrometer~\cite{Aam11} consists of a 3 T$\cdot$m dipole magnet, five tracking stations, each one based on two Cathode Pad Chambers, and two triggering stations, each one equipped with two planes of Resistive Plate Chambers. Two absorbers efficiently filter out hadrons. The front absorber, which is placed between
the interaction region and the muon spectrometer, has a thickness of 10 interaction lengths ($\lambda_{\rm I}$), while a second absorber, placed between the tracking and the triggering stations, has a thickness of 7.2 $\lambda_{\rm I}$. 
The other detectors used in this analysis are the two Si pixel layers corresponding to the 
innermost sections of the Inner Tracking System (ITS) ($|\eta_{\rm lab}|<0.9$){\cite{Aam10}, for the determination of the interaction vertex, and the two VZERO scintillator hodoscopes 
($2.8< \eta_{\rm lab} < 5.1$ and $-3.7 < \eta_{\rm lab} < -1.7$)~\cite{Abb13}, mainly for triggering purposes and for removing beam-induced background. The Zero Degree Calorimeters (ZDC)~\cite{Abe12b}, positioned symmetrically at 112.5 m from the interaction point, are used to clean the event sample by removing de-bunched proton-lead collisions. More details on the ALICE experimental setup can be found elsewhere~\cite{Aam08}.

Minimum-bias (MB) events are triggered requiring the coincidence of a signal in the two VZERO detectors. The efficiency of such a trigger for selecting non single-diffractive collisions is $>99$\%~\cite{Abe13b}.
A simulation based on Monte-Carlo (MC) event generators has shown that the contamination from single-diffractive and electromagnetic interactions is negligible~\cite{Abe13b}.
 Opposite-sign muon pairs are selected by means of a dimuon trigger given by the coincidence of a MB trigger with the detection of two muon candidate tracks in the trigger system of the muon spectrometer. The dimuon trigger is configured in order to select muons having a transverse momentum $p_{{\rm T},\mu}>$0.5 GeV/$c$. The effect of this threshold is not sharp and the single-muon trigger efficiency reaches its plateau value ($\sim 96$\%) for $p_{{\rm T},\mu}\sim$1.5 GeV/$c$. 
Events with more than a single interaction per bunch crossing (pile-up events) 
represent $\sim$2\% of MB triggered events, while the probability of having two dimuon triggers in the same bunch crossing is negligible.

Due to the energy asymmetry of the LHC beams ($E_{\rm p}=4$ TeV, $E_{\rm Pb}=1.58\cdot A_{\rm Pb}$ TeV, where $A_{\rm Pb}$=208 is the Pb-nucleus mass number) the nucleon-nucleon center-of-mass system of the collisions does not coincide with the laboratory system, but is shifted by $\Delta y=0.465$ in the direction of the proton beam. Data have been taken with 
two beam configurations, by inverting the sense of the orbits of the two particle species. In this way the regions 
$2.03<y_{\rm cms}<3.53$ and $-4.46<y_{\rm cms}<-2.96$ have been studied, where positive rapidities refer to the situation where the proton beam is 
travelling towards the muon spectrometer (in the following these configurations are referred to as \mbox{p-Pb} and \mbox{Pb-p}, respectively). 
The integrated luminosities used in this analysis for the two configurations are $5.01\pm 0.17$ nb$^{-1}$ (\mbox{p-Pb}) and $5.81\pm 0.18$ nb$^{-1}$ (\mbox{Pb-p}). These values
are determined using $\sigma^{\rm MB}_{\rm pPb}$ = $2.09\pm 0.06$ b and $\sigma^{\rm MB}_{\rm Pbp}$ = $2.12\pm 0.06$ b, estimated by means of van-der-Meer scans of the MB trigger signal~\cite{Gag13}.

An offline selection is performed in order to reject beam-induced background by requiring the signal timing in the VZERO and ZDC to be compatible with that of a nominal \mbox{p-Pb} interaction.
Candidate muon tracks are reconstructed in the muon tracking chambers using the standard reconstruction algorithm~\cite{Aam11}. It is then required that the two reconstructed tracks match a track segment in the trigger chambers (trigger tracklet). A further selection cut is applied  by requiring the muon tracks to exit the front absorber at a
radial distance from the beam axis $17.6<R_{\rm abs}<89.5$ cm, in order to reject tracks crossing its high-density section, where multiple scattering and energy loss effects are large. 
The condition $-4<\eta_{{\rm lab},\mu} < -2.5$ for single muons is required, in order to reject muons at the edge of the spectrometer's acceptance. No cut on the $z$-position of the interaction vertex is carried out, since it was verified that the dimuon invariant mass resolution does not depend on it. 

The extraction of the number of J/$\psi$ is performed starting from the invariant mass distributions of opposite sign
muon pairs in the kinematic domain $2.5<|y_{\rm lab}|<4$, $p_{\rm T}<15$ GeV/$c$, shown in Fig.~\ref{fig:1}. The distributions are fitted by means of a superposition of a continuum and a resonance shape. The continuum is parameterized either as a 
polynomial times an exponential function or as a Gaussian with a width linearly varying with mass, while for the resonance either a
Crystal Ball function~\cite{Gai82} with asymmetric tails at both high and low mass was chosen, or various pseudo-Gaussian
functions (see, for example, \cite{Sha01}). The non-Gaussian tails of the resonance shapes
are fixed in the fit to the values obtained in a MC simulation of the J/$\psi$ signal, while the mean and the width of
the Gaussian core are left as free parameters, since the large signal-over-background ratio ($\sim 2$ at 3-$\sigma$ level) allows the
data themselves to better constrain these parameters. The widths extracted from data ($\sim 70$ MeV/$c^{2}$) are larger by $\sim 10$\% than those 
from MC, and the mean value of the J/$\psi$ mass coincides with the nominal PDG value~\cite{Ber12} within $\sim 0.1$\%. 
The $\psi(2S)\rightarrow\mu^+\mu^-$ decay is taken into account in the fit function, but its effect on the determination of the 
number of J/$\psi$ events is negligible. The latter number is obtained as an average of the integral of the signal function over the
various fits, and the systematic uncertainty on this quantity is taken as the 1-$\sigma$ spread of the number of signal events. The result is  
$N_{\rm {pPb}}^{\rm J/\psi} = (6.69 \pm 0.05 \pm 0.08)\cdot 10^{4}$ and 
$N_{\rm {Pbp}}^{\rm J/\psi} = (5.67 \pm 0.05 \pm 0.07)\cdot 10^{4}$, where the first uncertainty is statistical and the second is systematic.

\begin{figure}[htbp]
\centering
\resizebox{0.48\textwidth}{!}
{\includegraphics*[bb=0 0 564 411]{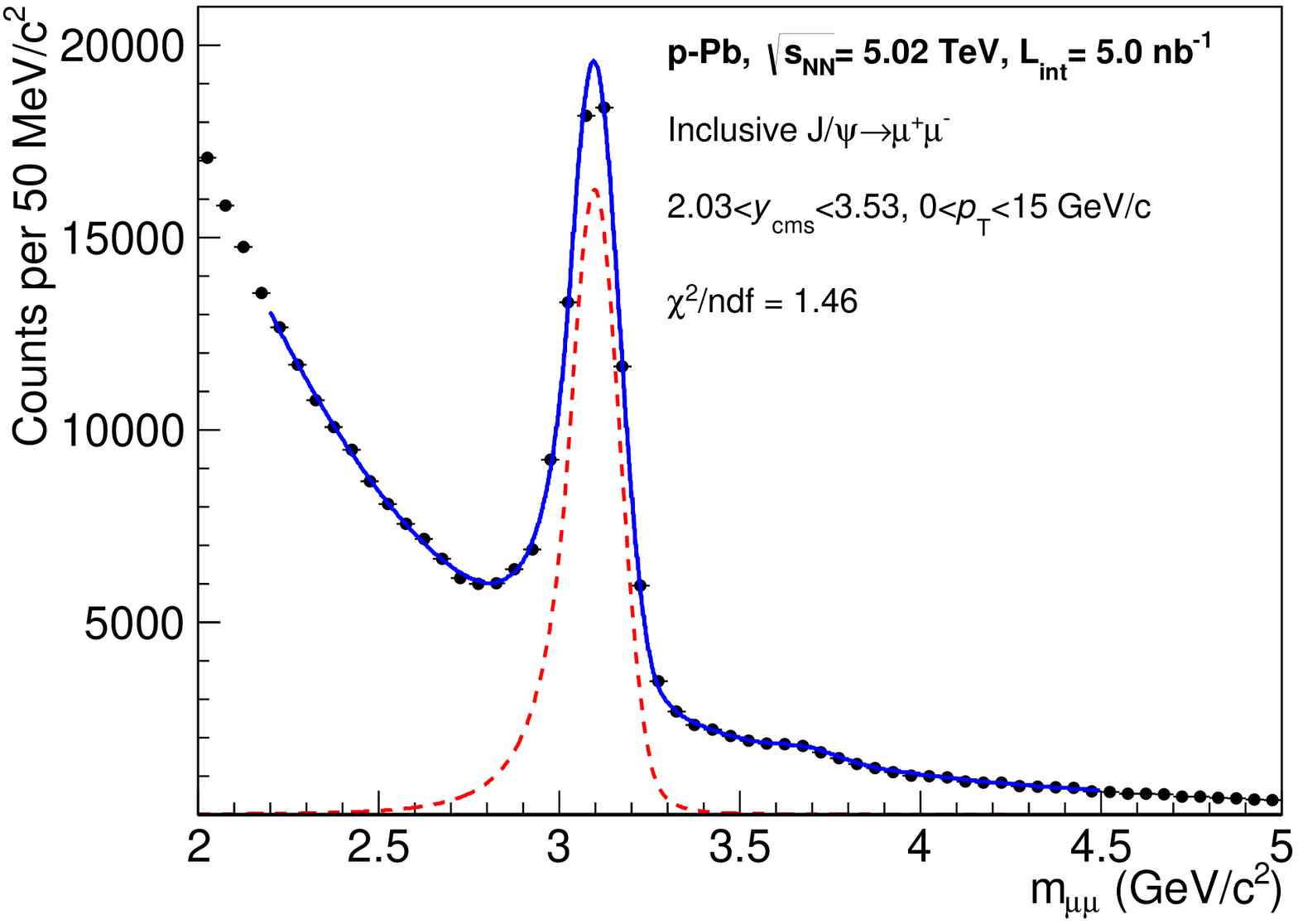}}
\resizebox{0.48\textwidth}{!}
{\includegraphics*[bb=0 0 564 411]{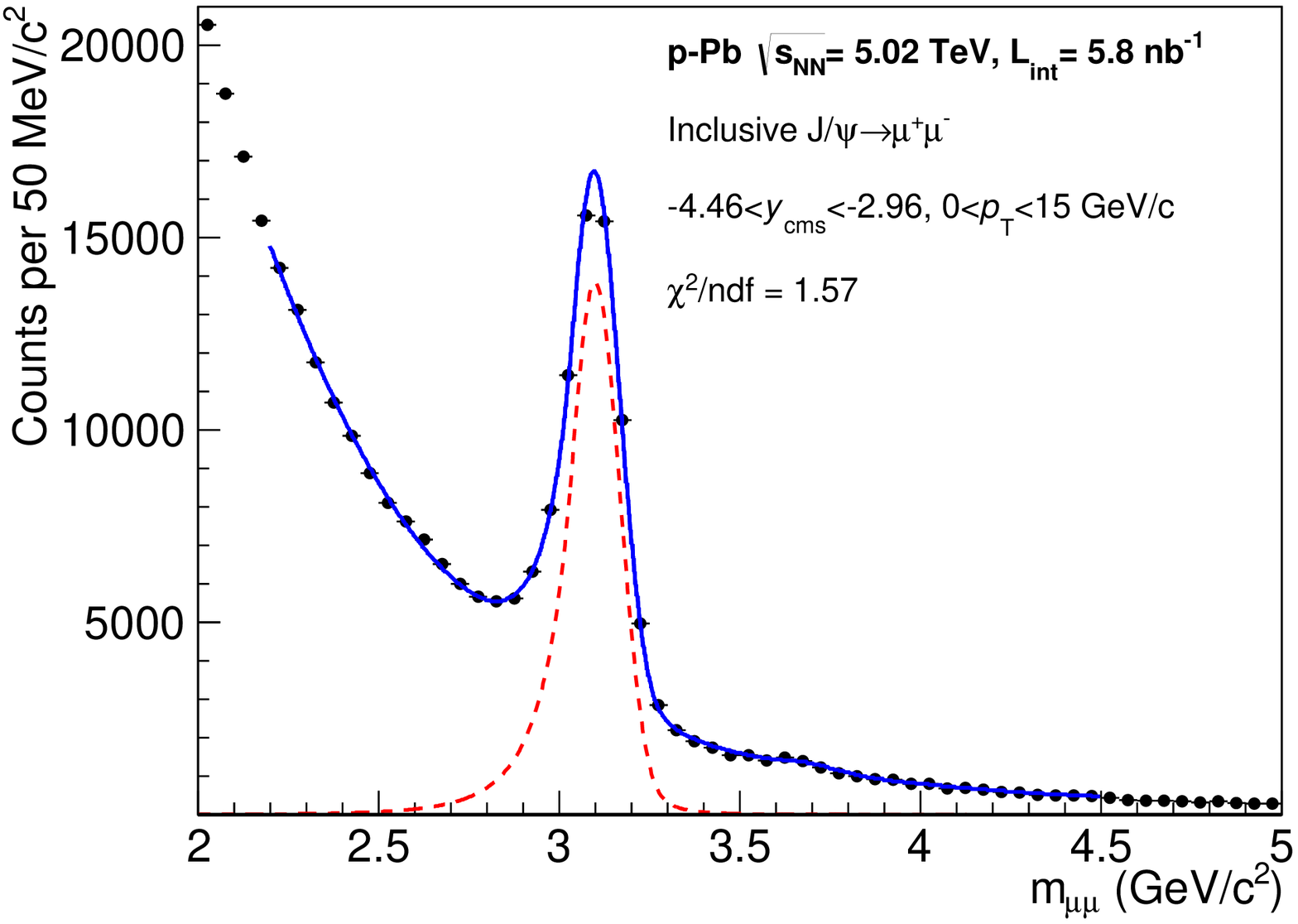}}
\caption{The opposite-sign dimuon invariant mass spectra for the \mbox{p-Pb} (left) and \mbox{Pb-p} (right) data samples, together with the result of the fit. For the 
fits shown here a Crystal Ball function (shown as a dashed line) and a variable-width Gaussian have been used for the signal and the background, respectively.}
\label{fig:1}
\end{figure}

The number of measured J/$\psi$ is then divided by the product of acceptance times efficiency $A\cdot\epsilon$, which is obtained using a MC 
simulation of the J/$\psi$ signal. An unpolarized distribution for the J/$\psi$ is assumed, following the small degree of
polarization measured in pp collisions at $\sqrt{s}=7$ TeV~\cite{Abe12p,Aai13,Cha13}, while the $p_{\rm T}$ and $y$ distributions used as an input for the generator are tuned to the measured data through an iterative procedure. 
The systematic uncertainty on the acceptance is obtained by defining $y$ ($p_{\rm T}$) distributions for selected phase space regions,
corresponding to sub-ranges in $p_{\rm T}$ ($y$) and centrality of the collision. The hardest and softest spectra for each variable are then used as inputs to the MC calculation, and the variation with respect to the default acceptance values gives the systematic uncertainty, which amounts to 1.5\% for both \mbox{p-Pb} and {Pb-p}.

The efficiency of the muon triggering detectors is calculated with a procedure based on data and involving the analysis of
trigger tracklets constructed from hits in the four planes of the two trigger stations. For the tracking chambers a map of dead channels is obtained from the online detector information and updated on a run-per-run basis. Both information are injected in the MC and their time evolution is taken into account by performing a simulation for each run, with a number of J/$\psi$ signal events proportional to the number of offline-selected triggered events. 

The systematic uncertainty on the J/$\psi$ trigger efficiency ($\epsilon_{\rm trig}^{\rm J/\psi}$) is obtained as a convolution of various sources. First, effects related to the estimate of the efficiency of the trigger detectors are studied by varying their efficiency in the MC by an amount equal to the statistical uncertainty on their 
evaluation ($\sim 2$\%). This results in a 2\% change in $\epsilon_{\rm trig}^{\rm J/\psi}$. Second, systematic effects related to small discrepancies in the $p_{\rm T}$ dependence of the muon trigger threshold between data and MC give a $\sim 2-2.5$\% contribution to $\epsilon_{\rm trig}^{\rm J/\psi}$. Finally, there is a $\sim 1$\% effect related to the choice of the goodness-of-fit $\chi^2$ cut used in defining the matching between tracking and triggering information.

The single-muon tracking efficiencies are obtained using an algorithm based on reconstructed tracks~\cite{Aam11}. The systematic uncertainty on this quantity is obtained by comparing the results obtained with MC and real data.
This uncertainty is considered as fully uncorrelated between the two detected muons and, at the dimuon level, it amounts to 4\% (6\%) for \mbox{p-Pb} (\mbox{Pb-p}). In addition, it was checked that the tracking efficiency does not depend on the centrality of the collision, justifying the use of pure signal MC simulations to determine $A\cdot\epsilon$.

The average $A\cdot\epsilon$ values for the two kinematic regions are $(25.4 \pm 1.3)$\% (\mbox{p-Pb}) and $(17.1 \pm 1.2)$\% (\mbox{Pb-p}). The quoted uncertainty is systematic, and the lower value for \mbox{Pb-p} is mainly due to a smaller detector efficiency in the corresponding data taking period.

The inclusive J/$\psi$ production cross section is 

\begin{equation}
\sigma^{\rm J/\psi}_{\rm pPb}=\frac{N^{\rm cor}_{\rm J/\psi\rightarrow\mu\mu}}
{N_{\rm MB} \cdot {\rm B.R.}({\rm J}/\psi\rightarrow\mu\mu)}
\times{\sigma^{\rm MB}_{\rm pPb}}
\end{equation}

where ${N^{\rm cor}_{\rm J/\psi\rightarrow\mu\mu}}$ is the number of J/$\psi$ corrected for $A\cdot\epsilon$, 
${\rm B.R.}({\rm J}/\psi\rightarrow\mu\mu)=(5.93\pm 0.06)$\% is the branching ratio for the J/$\psi$ decay to 
dimuons~\cite{Ber12}, $N_{\rm MB}$ is the number of MB \mbox{p-Pb} collisions, and $\sigma^{\rm MB}_{\rm pPb}$ the corresponding 
cross section. 

Since the analysis is based on a dimuon trigger sample, the equivalent number of MB triggers is evaluated as $F\cdot N_{\rm DIMU}$, where $N_{\rm DIMU}$ is the number of opposite sign dimuon triggered events, which amounts to $9.27\cdot 10^6$ for \mbox{p-Pb} and $2.09\cdot 10^7$ for \mbox{Pb-p}. The enhancement factor $F$ is calculated in two different ways. In the first one
it is obtained as the product $F_{2\mu/1\mu}\cdot F_{1\mu/MB}$, where $F_{2\mu/1\mu}$ is the inverse of the probability of having a second muon triggered when one muon has triggered the event and, correspondingly, $F_{1\mu/MB}$ is the inverse of the probability of having one triggered muon in events where the MB condition is required. The various quantities are obtained from the recorded trigger mask for the collected events after quality cuts. Obtaining $F$ as the product of the two factors mentioned above allows the statistical uncertainty to be reduced. In the second approach, the information of the counters recording the number of level-0 triggers is used. In this case, statistics are much larger and $F$ is obtained as the ratio between the numbers of MB and dimuon triggers at level-0, corrected for pile-up effects (2\%) and taking into account the slight difference in the fraction of events surviving the quality cuts for the two trigger samples (1\%). One gets, averaging the results from the two approaches, 
$F_{\rm pPb}=1129 \pm 2$ and $F_{\rm Pbp}=589 \pm 2$, where the quoted uncertainties are statistical. A 1\% systematic uncertainty is estimated on both quantities, corresponding to the difference between the values obtained in the two calculations. 

Finally, the quantity $N_{\rm MB}/\sigma^{\rm MB}_{\rm pPb}$ corresponds to the integrated luminosity. 
As a cross-check, its value has been measured independently by using a second reference trigger, issued by a \v{C}herenkov counter~\cite{Aam08}, whose cross section was also measured in the van-der-Meer scans. The luminosities measured with the two luminometers differ by at most 1\% throughout the whole data-taking period. This small difference (identical for \mbox{p-Pb} and \mbox{Pb-p}) has been included in the systematic uncertainty on $\sigma^{\rm MB}_{\rm pPb}$.

The resulting cross sections are 
\begin{eqnarray*}
\sigma^{\rm J/\psi}_{\rm pPb} (2.03<y_{\rm cms}<3.53) = 886 \pm 6 {\rm (stat.)} \pm 48 {\rm (syst. uncorr.)} \pm 30 {\rm (syst. part. corr.)}\,\, \mu b \\
\sigma^{\rm J/\psi}_{\rm Pbp}(-4.46<y_{\rm cms}<-2.96) = 966 \pm 8 {\rm (stat.)} \pm 70 {\rm (syst. uncorr.)} \pm 31 {\rm (syst. part. corr.)}\,\, \mu b 
\end{eqnarray*}

The uncertainties connected with tracking, matching and triggering efficiency, with signal extraction, with the choice of the MC input distributions and with the evaluation of $N_{\rm MB}$ are taken as uncorrelated between \mbox{p-Pb} and \mbox{Pb-p}, while those on $\sigma^{\rm MB}$ are partially correlated. In the latter uncertainty a 1\% contribution due to the uncertainty on B.R.(${\rm J}/\psi\rightarrow\mu\mu$) was also included. A summary of the sources of sytematic uncertainties and their numerical values are given in Table~\ref{tab:1}.

\begin{table}
\caption{\label{tab:1} Systematic uncertainties (in percent) contributing to the measurement of inclusive J/$\psi$ cross sections and nuclear modification factors. When the uncertainty values depend on the rapidity bin under consideration, their maximum and minimum values are quoted. Uncertainties on 
$\sigma^{\rm MB}_{\rm pPb}$ are relevant for inclusive J/$\psi$ cross sections only, while those on $\sigma^{{\rm J/}\psi}_{\rm pp}$ and $\langle T_{\rm pPb}\rangle$ contribute only to the uncertainty on the nuclear modification factors.}

\centering
\begin{tabular}{c|c|c}
\hline
\hline
Source & $\sigma^{{\rm J/}\psi}_{\rm pPb}$, $R_{\rm pPb}$ & $\sigma^{{\rm J/}\psi}_{\rm Pbp}$, $R_{\rm Pbp}$ \\ \hline
{\it Uncorrelated} & & \\
Tracking efficiency & 4 & 6 \\
Trigger efficiency & 2.8 & 3.2 \\
Signal extraction & 1.3 (1.5 $-$ 3.4) & 1.2 (1.6 $-$ 3.8) \\
MC input & 1.5 (1.1 $-$ 3) & 1.5 (0.9 $-$ 4.2)\\
Matching efficiency & 1 & 1 \\ 
$F$ & 1 & 1 \\ 
$\sigma^{{\rm J/}\psi}_{\rm pp}$ 
& 4.3 (3.1 $-$ 6.0) & 4.6 (3.1 $-$ 13.4) \\ \hline
{\it Partially correlated} & & \\
$\sigma^{\rm MB}_{\rm pPb}$ 
& 3.2 & 3 \\ 
$\sigma^{{\rm J/}\psi}_{\rm pp}$ 
& 3.7 (2.7 $-$ 9.2) & 3.1 (1.2 $-$ 8.3) \\ \hline
{\it Correlated} & \multicolumn {2}{c} {}  \\
B.R. & \multicolumn {2}{c} {1}  \\
$\langle T_{\rm pPb}\rangle$ 
& \multicolumn{2}{c} {3.6} \\ 
$\sigma^{{\rm J/}\psi}_{\rm pp}$ 
& \multicolumn {2}{c} {5.5}  \\ \hline
\end{tabular}
\end{table}

The nuclear effects on J/$\psi$ production are quantified using the nuclear modification factor $R_{\rm pPb}$, obtained as
\begin{equation}
R_{\rm pPb}=\frac{N^{\rm cor}_{\rm J/\psi\rightarrow\mu\mu}}{\langle T_{\rm pPb}\rangle \cdot
N_{\rm MB} \cdot {\rm B.R.}({\rm J}/\psi\rightarrow\mu\mu) \cdot \sigma^{\rm J/\psi}_{\rm pp}}
\end{equation}

where $\sigma^{\rm J/\psi}_{\rm pp}$ is the production cross section in pp collisions in the same kinematical domain and at the same $\sqrt{s}$ (the same formula applies to \mbox{Pb-p}), and 
$\langle T_{\rm pPb}\rangle$ is the nuclear thickness function estimated through the Glauber model, which gives $\langle T_{\rm pPb}\rangle = 0.0983 \pm 0.0035$ mb$^{-1}$~\cite{Abe13}. The uncertainty on $\langle T_{\rm pPb}\rangle$ was obtained by varying the parameters of the Glauber model.

Since pp data at $\sqrt{s}=5.02$ TeV are not available, the reference cross section $\sigma^{\rm J/\psi}_{\rm pp}$ has been obtained by means of an interpolation procedure~\cite{ALPN}, based on forward rapidity ($2.5<y_{\rm cms}<4$) pp results at $\sqrt{s}=2.76$ and 7 TeV from ALICE~\cite{Abe12c,Abe13p}. 
The $\sqrt{s}$-interpolation is based on three empirical shapes (linear, power law, exponential) and is independently performed for each of the six rapidity bins corresponding to the ${\rm d}\sigma/{\rm d}y$ values measured at the two energies. The central values of the interpolation are given, for each rapidity bin, by the average of the three values obtained with the adopted shapes. Their uncertainties are the quadratic sum of a dominant term, related to the uncertainties on the points used for the interpolation, and of a term corresponding to the maximum spread between the results obtained with the various shapes. 
A small additional systematic uncertainty is obtained comparing the empirical shapes with those calculated with the leading order (LO) CEM~\cite{Glu78} and FONLL~\cite{Cac98} models. We recall that the CEM (Color Evaporation Model) assumes that a fixed fraction of $c\overline{c}$ pairs produced with an invariant mass $m<2m_{\rm D}$ ends up in producing charmonium states. Although it does not contain a dynamical description of the production process, it was shown to be phenomenologically successful over a large $\sqrt{s}$ range. FONLL gives predictions for the total $c\overline c$ production rather than for the J/$\psi$ cross section, but we assume, similarly to the CEM approach, that the fraction of $c\overline c$ pairs going to charmonium is $\sqrt{s}$-independent.

Due to the $\Delta y=0.465$ rapidity shift induced by the asymmetry in the energy per nucleon of the proton and lead beams, the rapidity regions covered by
the present analysis do not correspond to the ones available for pp. Therefore, the ${\rm d}\sigma/{\rm d}y$ values obtained at $\sqrt{s}=5.02$ TeV with the
procedure described above have been fitted to various shapes (Gaussian, second and fourth order polynomials~\cite{Bos11}). The values for $\sigma^{\rm J/\psi}_{\rm pp}$ at $\sqrt{s}=5.02$ TeV for the \mbox{p-Pb} and \mbox{Pb-p} rapidity intervals were finally obtained as the average of the integral of the various fitting functions in the corresponding $y$-ranges, and are ${\rm B.R.}\cdot \sigma^{\rm J/\psi}_{\rm pp}(2.03<y_{\rm cms}<3.53)= 367 \pm 29$ nb and ${\rm B.R.}\cdot \sigma^{\rm J/\psi}_{\rm pp}(-4.46<y_{\rm cms}<-2.96) = 255 \pm 20 $ nb~\cite{ALPN}. The quoted total uncertainties include again a contribution from the maximum spread of the results obtained with the various functions. 

The measured nuclear modification factors, shown in Fig.~\ref{fig:2}, are 
\begin{eqnarray*}
R_{\rm pPb}(2.03 < y_{\rm cms} < 3.53) = 0.70 \pm 0.01 {\rm (stat.)} \pm 0.05 {\rm (syst. uncorr.)} \pm 0.03 {\rm (syst. part. corr.)} \pm 0.05 {\rm (syst. corr.)} \;\;\;\;\;\;\; \\
R_{\rm Pbp}(-4.46 < y_{\rm cms} < -2.96) = 1.08 \pm 0.01 {\rm (stat.)} \pm
0.09 {\rm (syst. uncorr.)} \pm 0.03 {\rm (syst. part. corr.)} \pm 0.07 {\rm (syst. corr.)}
\end{eqnarray*}

At forward rapidity the inclusive J/$\psi$
production is suppressed with respect to the one in binary-scaled pp collisions, whereas it is unchanged at backward rapidity. The uncertainties related to $\langle T_{\rm pPb}\rangle$ and B.R.(${\rm J}/\psi\rightarrow\mu\mu$) are  considered as correlated. The uncertainties connected with tracking, matching and triggering efficiencies, with signal extraction, with the choice of the MC input distributions, and with the evaluation of $N_{\rm MB}$ are taken as uncorrelated. Finally, the uncertainty on the pp cross section interpolation is splitted (see~\cite{ALPN} for details) among the three uncertainties quoted for the nuclear modification factors. The numerical details on systematic uncertainties are given in Table~\ref{tab:1}. Our measurements are compared with a next to leading order (NLO) CEM calculation which uses the EPS09 shadowing parameterization~\cite{Vog13}, and with the result of a theoretical prediction which includes a contribution from coherent parton energy loss processes~\cite{Arl12b}, either in addition to EPS09 shadowing or as the only nuclear effect. Finally, results from a calculation in the CGC framework~\cite{Fuj13}, combined with a CEM production model, are also shown. Within our uncertainties, both the model based on shadowing only and the coherent energy loss approach are able to describe the data, while the CGC-based prediction overestimates the observed suppression. None of these models include a suppression related to the break-up of the $c\overline{c}$ pair.

It is worth noting that calculations refer to prompt production (i.e., direct J/$\psi$ plus the contribution
from $\psi$(2S) and $\chi_{\rm c}$ decays), while the experimental results are for inclusive J/$\psi$ production, which contains 
a non-prompt contribution from B-decays. However, the $p_{\rm T}$-integrated non-prompt J/$\psi$ fraction is small (LHCb measured 7.1\% at $\sqrt{s}=2.76$ TeV in the kinematic region $2<y_{\rm cms}<4.5$, $p_{\rm T}<12$ GeV/$c$~\cite{Aai13cs} and 9.8\% at $\sqrt{s}=7$ TeV for $2<y_{\rm cms}<4.5$, $p_{\rm T}<14$ GeV/$c$~\cite{Aai11}). The difference between $R_{\rm pPb}^{\rm incl}$ and $R_{\rm pPb}^{\rm prompt}$ is well within the uncertainties of our measurement for a very large range of $R_{\rm pPb}^{\rm non-prompt}$, from almost complete suppression ($R_{\rm pPb}^{\rm non-prompt}=0.2$) to a rather  strong enhancement ($R_{\rm pPb}^{\rm non-prompt}=1.3$). 
A similar conclusion holds at backward rapidity.

\begin{figure}[htbp]
\centering
\resizebox{0.7\textwidth}{!}
{\includegraphics*[bb=0 0 564 411]{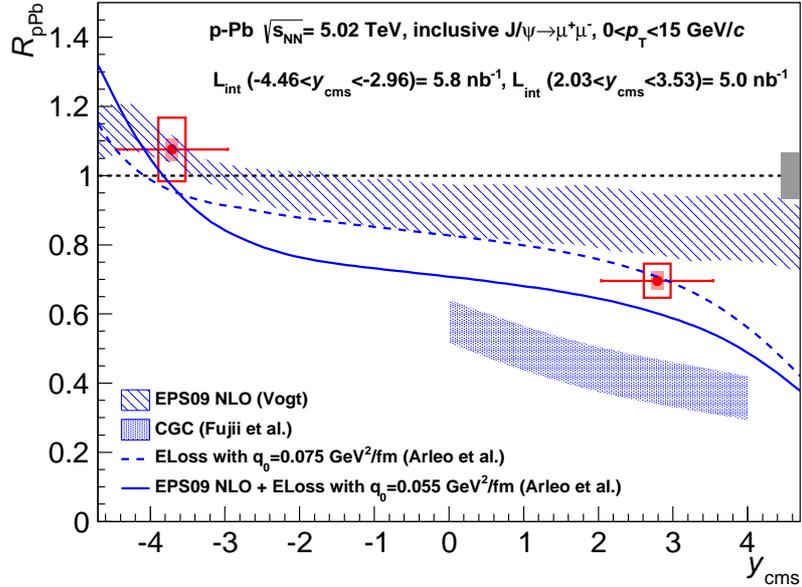}}
\caption{The nuclear modification factors for inclusive J/$\psi$ production at $\sqrt{s_{\rm NN}}=5.02$ TeV. The error bars correspond to the statistical uncertainties, the open boxes to the uncorrelated systematic uncertainties, the shaded boxes around the points represent the partially correlated systematic uncertainties. 
The box around $R_{\rm pPb}=1$ shows the size of the correlated uncertainties.
Results from various models are also shown. The theoretical uncertainties for the EPS09 NLO calculation~\cite{Vog13} are due to the uncertainty on the EPS09 shadowing parameterization and to the mass and scale uncertainties on the cross section calculation. For the CGC model~\cite{Fuj13}, the band is related to the choice of the parton saturation scale and of the charm quark mass. Finally, the $q_{\rm 0}$ value in the energy loss model~\cite{Arl12b} represents the value of the transport coefficient in the target nucleons for $x_{\rm Bj}$=10$^{-2}$ gluons. 
}
\label{fig:2}
\end{figure}

Both cross sections and nuclear modification factors for inclusive J/$\psi$ have also been studied differentially in rapidity, with six bins for each of the two $y$ domains. The results are shown in Fig.~\ref{fig:3a} and Fig.~\ref{fig:3}, respectively.
The analysis procedure is identical to the one detailed above for the study of the integrated quantities. In particular, a differential estimate of the systematic uncertainties for the various ingredients has been carried out.  
The larger uncertainties visible at the lower edges of the rapidity ranges covered in \mbox{p-Pb} and \mbox{Pb-p} are related to a larger uncertainty on the pp reference cross section, due to the fact that these regions are not directly covered by the pp measurements and therefore an extrapolation has to be performed~\cite{ALPN}.
No strong variation of the nuclear modification factors is observed, in particular at backward rapidity, where models including coherent energy loss suggest a steeper behaviour.

\begin{figure}[htbp]
\centering
\resizebox{0.7\textwidth}{!}
{\includegraphics*[bb=0 0 564 411]{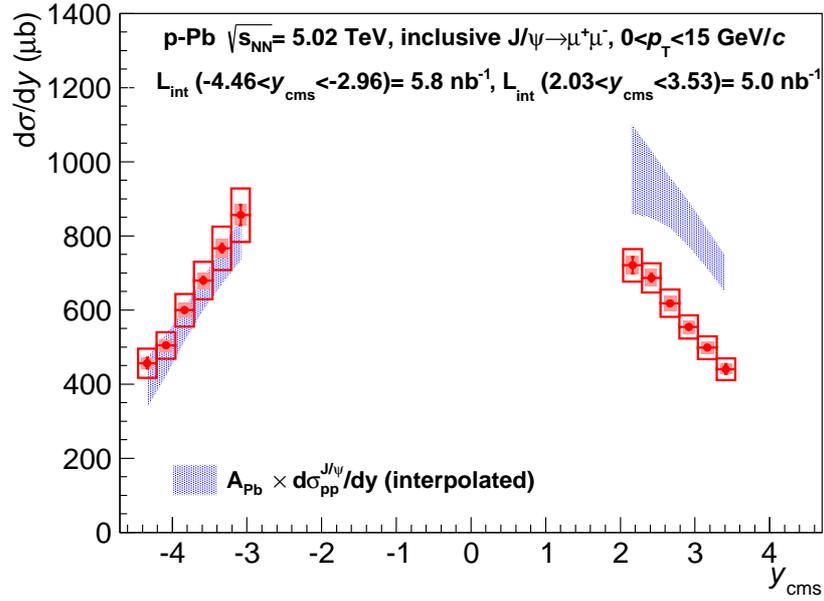}}
\caption{The inclusive J/$\psi$ production cross section, as a function of rapidity. The error bars correspond to the statistical uncertainties, the open boxes to the uncorrelated systematic uncertainties, the shaded boxes around the points represent the partially correlated systematic uncertainties. 
The bands correspond to the inclusive J/$\psi$ pp cross section, obtained with the interpolation procedure described in the text and scaled by the Pb-nucleus mass number $A_{\rm Pb}$.}
\label{fig:3a}
\end{figure}

\begin{figure}[htbp]
\centering
\resizebox{0.7\textwidth}{!}
{\includegraphics*[bb=0 0 564 411]{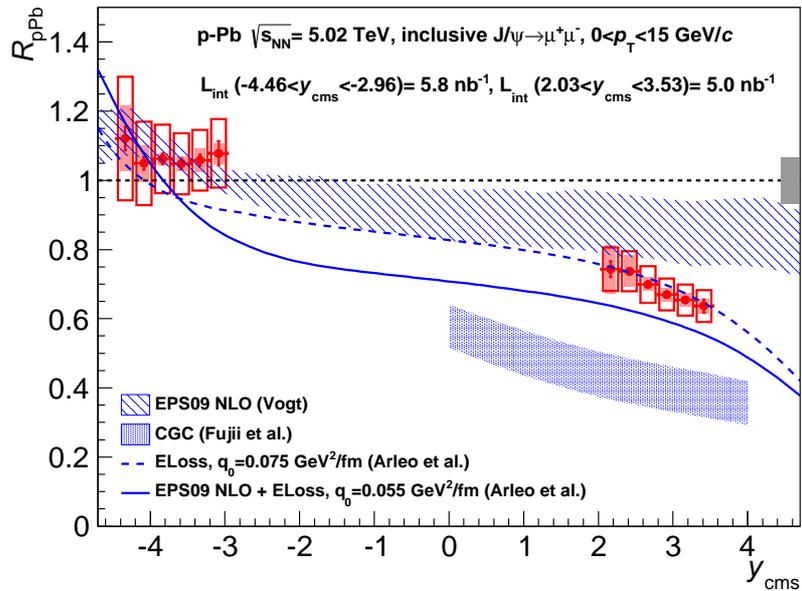}}
\caption{The nuclear modification factors for inclusive J/$\psi$ production at $\sqrt{s_{\rm NN}}=5.02$ TeV, in bins of rapidity. The meaning of symbols and curves is the same as in Fig.~\ref{fig:2}.}
\label{fig:3}
\end{figure}

Both $\sigma^{\rm J/\psi}_{\rm pp}$ 
and $\langle T_{\rm pPb}\rangle$ cancel out when forming the ratio $R_{\rm FB}$ of the nuclear modification factors for a rapidity range symmetric with respect to $y_{\rm cms}=0$. In this way one is left with the ratio of the forward and backward J/$\psi$ yields. The drawback of this approach is that, due to the beam energy asymmetry, the common $y$ interval covered at both forward and backward rapidity is smaller than the acceptance of the muon spectrometer, and limited to $2.96<|y_{\rm cms}|<3.53$. The reduction in statistics by a factor $\sim$3 is compensated by the cancellation of the reference-related uncertainties. 
The obtained value is $R_{\rm FB}(2.96<|y_{\rm cms}|<3.53)= 0.60 \pm 0.01 {\rm (stat.)} \pm 0.06 {\rm (syst.)}$. The systematic uncertainties which are uncorrelated between backward and forward rapidity (tracking, matching and triggering efficiency, normalization, MC input) have been quadratically combined in the ratio, while for signal extraction the uncertainty has directly been calculated on the ratio of the number of J/$\psi$. The main contribution to the $R_{\rm FB}$ uncertainty comes from the tracking efficiency.

In Fig.~\ref{fig:4} we show a comparison of $R_{\rm FB}$ with the results of the theoretical calculations discussed above, except for the CGC-inspired model, which gives predictions only at forward rapidity. In addition, a prediction based on a LO approach, implementing a $2\rightarrow 2$ kinematics ($gg\rightarrow J/\psi\,g$) and using either the EPS09 or the nDSG shadowing parameterization, is also shown~\cite{Lan13}. The agreement between data and the model including both shadowing and coherent energy loss is very good, while pure shadowing scenarios seem to overestimate $R_{\rm FB}$.
However, it has to be noted that, although the experimental measurement of $R_{\rm FB}$ has a smaller uncertainty than $R_{\rm pPb}$
and $R_{\rm Pbp}$, its comparison with theoretical calculations is less stringent, since models which globally overestimate/underestimate the nuclear modification factors may still provide a very good agreement with the measured $R_{\rm FB}$.

\begin{figure}[htbp]
\centering
\resizebox{0.7\textwidth}{!}
{\includegraphics*[bb=0 0 564 411]{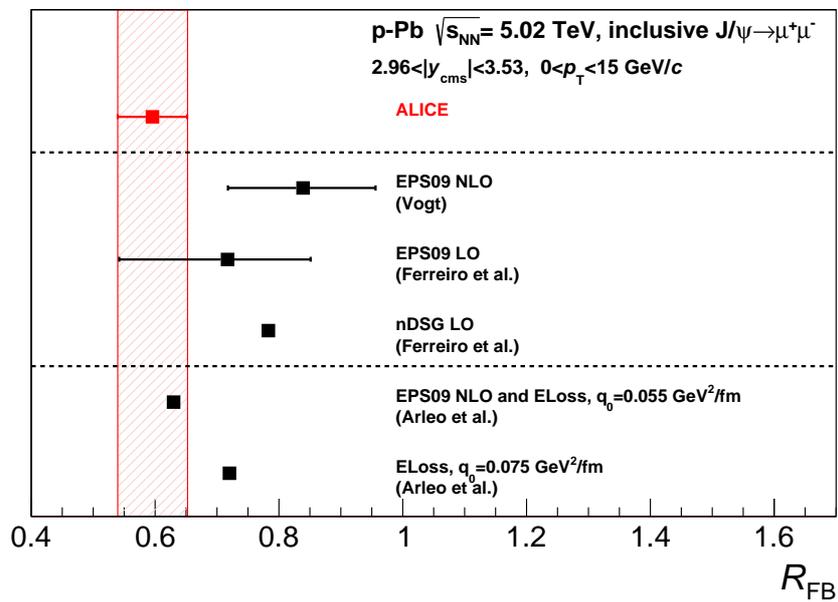}}
\caption{The forward to backward ratio $R_{\rm FB}$ of the nuclear modification factors for inclusive J/$\psi$ production, compared to theoretical models. The statistical and systematic uncertainties for the experimental value are added in quadrature. For the shadowing calculations, uncertainties are quoted when available, and are obtained in the same way as in Fig.~\ref{fig:2}.}
\label{fig:4}
\end{figure}

The $R_{\rm FB}$ ratio has also been studied differentially in $y$ (3 bins) and
$p_{\rm T}$ (10 bins, covering the region $p_{\rm T}<$15 GeV/{\rm c}). In Fig.~\ref{fig:5} we show the results, again compared with the predictions of the models. The treatment of the uncertainties is the same described above for the integrated value of $R_{\rm FB}$. As a function of rapidity, no variation is observed in the relatively narrow region covered by the $R_{\rm FB}$ measurement, while a trend towards higher $R_{\rm FB}$ values is seen as $p_{\rm T}$ increases. Models including coherent energy loss seem to qualitatively reproduce the data, in particular when shadowing effects are taken into account, although they predict a steeper behaviour at low $p_{\rm T}$~\cite{Arl13}. 

\begin{figure}[htbp]
\centering
\resizebox{0.7\textwidth}{!}
{\includegraphics*[bb=0 0 564 411]{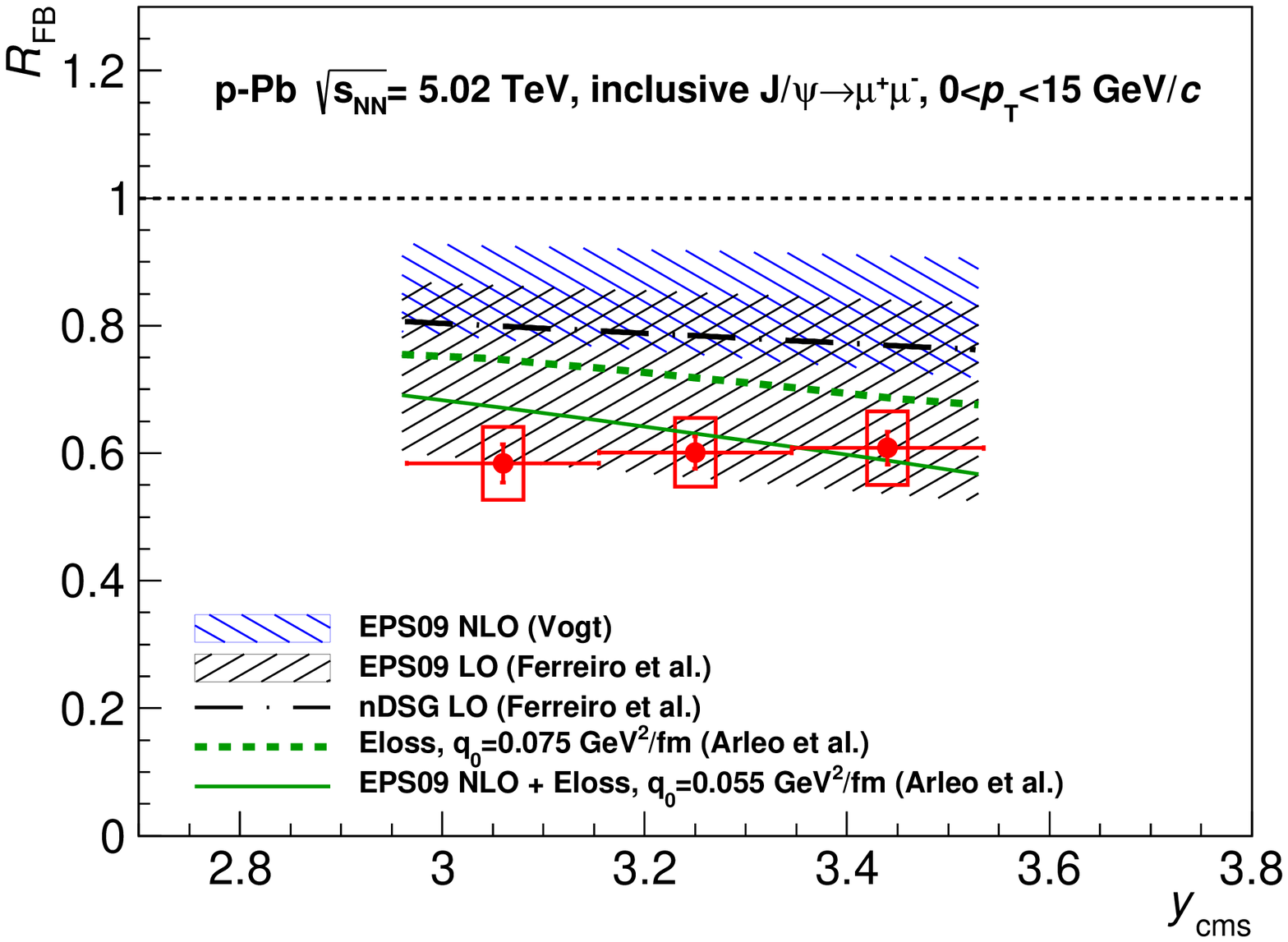}}
\resizebox{0.7\textwidth}{!}
{\includegraphics*[bb=0 0 564 408]{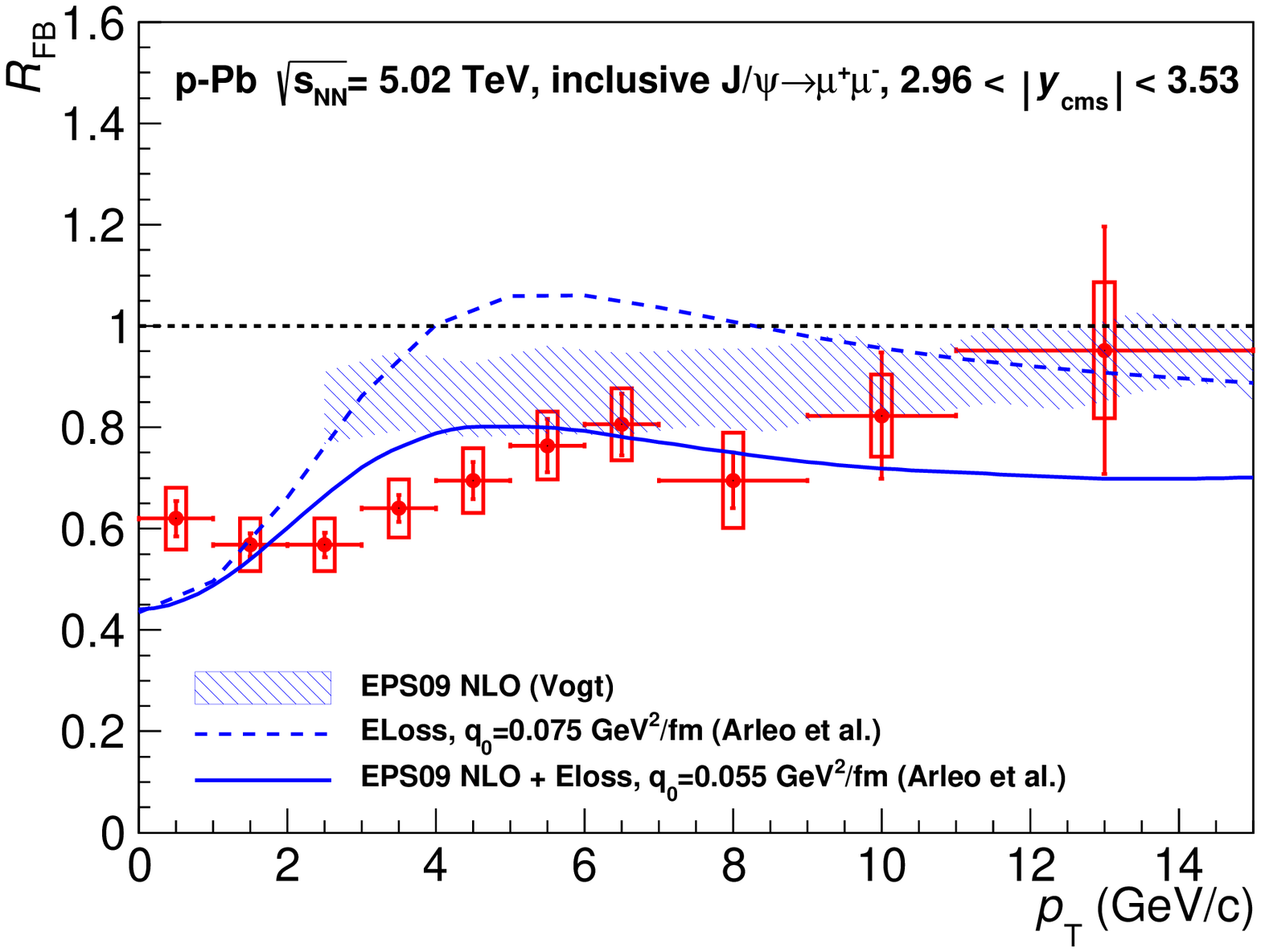}}
\caption{The forward to backward ratio $R_{\rm FB}$ of the nuclear modification factors for inclusive J/$\psi$ production, as a function of $y$ and $p_{\rm T}$, compared to theoretical models. The bars represent the statistical uncertainty, while the open boxes correspond to uncorrelated systematic uncertainties.}
\label{fig:5}
\end{figure}

Finally, the results presented in this paper provide information on the magnitude of cold nuclear matter effects in \mbox{Pb-Pb} collisions. ALICE has published results for $R_{\rm PbPb}$ in the region $2.5<y_{\rm cms}<4$ at $\sqrt{s_{\rm NN}}=2.76$ TeV~\cite{Abe12,Abe13J}. Although the \mbox{p-Pb} data discussed above refer to slightly different $y_{\rm cms}$ regions and to a larger center of mass energy, the Bjorken-$x$ regions probed by the J/$\psi$ production process in the Pb nuclei for \mbox{p-Pb} and \mbox{Pb-Pb} are shifted by $\sim$10\% only.  Indeed, in the so-called 2$\rightarrow$1 approach, where the production kinematics is $gg\rightarrow J/\psi$~\cite{Vog04}, the $x$-values selected in \mbox{Pb-Pb} collisions are $2\cdot 10^{-5}<x<9\cdot 10^{-5}$, $1\cdot 10^{-2}<x<6\cdot 10^{-2}$, the two ranges being relative to nucleons moving away from or toward the muon spectrometer, respectively. In proton-nucleus collisions, the probed $x$
values for nucleons inside the Pb nucleus are $2\cdot 10^{-5}<x<8\cdot 10^{-5}$ for \mbox{p-Pb} and $1\cdot 10^{-2}<x<5\cdot 10^{-2}$ for \mbox{Pb-p}. If shadowing is the main nuclear effect, a hypothesis in fair agreement with the results shown in this paper, as a first approximation cold nuclear matter effects on $R_{\rm PbPb}$ would be given by the product $R_{\rm pPb}\times R_{\rm Pbp}$. 
This product is $0.75 \pm 0.10 \pm 0.12$ (the first uncertainty being related 
to the quadratical combination of statistical and uncorrelated systematic uncertainties, the second one coming from the linear combination of correlated uncertainties), which is larger than 
$R_{\rm PbPb}=0.57\pm 0.01\pm 0.09$~\cite{Abe13J}. This could be an indication that the J/$\psi$ suppression effect observed in
\mbox{Pb-Pb} collisions cannot be ascribed to cold nuclear matter effects alone. However, the size of the uncertainties prevents a strong conclusion on this point.

In summary, inclusive J/$\psi$ production has been measured with the ALICE detector in \mbox{p-Pb} collisions at the CERN LHC. In this paper we have presented the production cross sections and the nuclear modification factors in the regions $2.03<y_{\rm cms}<3.53$ and $-4.46<y_{\rm cms}<-2.96$, as well as their ratio $R_{\rm FB}$ in the region $2.96<|y_{\rm cms}|<3.53$. While at forward rapidity ($R_{\rm pPb}$) a suppression with respect to pp collisions is observed, in the backward region ($R_{\rm Pbp}$) no suppression is present. A fair agreement is seen with predictions based on a pure nuclear shadowing scenario~\cite{Vog13,Lan13}, parameterized using the EPS09 approach, as well as with models including a contribution from coherent partonic energy loss~\cite{Arl12}. None of these models include a final
state break-up of the J/$\psi$ in cold nuclear matter. The study of $R_{\rm FB}$, carried out as a function of $y$ and $p_{\rm T}$, confirms these indications. 
Finally, the results presented in this paper provide an important baseline for the interpretation of heavy-ion collision results and 
are in agreement with those presented by the LHCb Collaboration~\cite{LHCb}.

%% file: acknowledgements_march2013.tex
The ALICE collaboration is grateful for useful discussions with the LHCb collaboration.
The ALICE collaboration would like to thank all its engineers and technicians for their invaluable contributions to the construction of the experiment and the CERN accelerator teams for the outstanding performance of the LHC complex.
The ALICE collaboration acknowledges the following funding agencies for their support in building and
running the ALICE detector:
State Committee of Science,  World Federation of Scientists (WFS)
and Swiss Fonds Kidagan, Armenia,
Conselho Nacional de Desenvolvimento Cient\'{\i}fico e Tecnol\'{o}gico (CNPq), Financiadora de Estudos e Projetos (FINEP),
Funda\c{c}\~{a}o de Amparo \`{a} Pesquisa do Estado de S\~{a}o Paulo (FAPESP);
National Natural Science Foundation of China (NSFC), the Chinese Ministry of Education (CMOE)
and the Ministry of Science and Technology of China (MSTC);
Ministry of Education and Youth of the Czech Republic;
Danish Natural Science Research Council, the Carlsberg Foundation and the Danish National Research Foundation;
The European Research Council under the European Community's Seventh Framework Programme;
Helsinki Institute of Physics and the Academy of Finland;
French CNRS-IN2P3, the `Region Pays de Loire', `Region Alsace', `Region Auvergne' and CEA, France;
German BMBF and the Helmholtz Association;
General Secretariat for Research and Technology, Ministry of
Development, Greece;
Hungarian OTKA and National Office for Research and Technology (NKTH);
Department of Atomic Energy and Department of Science and Technology of the Government of India;
Istituto Nazionale di Fisica Nucleare (INFN) and Centro Fermi -
Museo Storico della Fisica e Centro Studi e Ricerche "Enrico
Fermi", Italy;
MEXT Grant-in-Aid for Specially Promoted Research, Ja\-pan;
Joint Institute for Nuclear Research, Dubna;
National Research Foundation of Korea (NRF);
CONACYT, DGAPA, M\'{e}xico, ALFA-EC and the EPLANET Program
(European Particle Physics Latin American Network)
Stichting voor Fundamenteel Onderzoek der Materie (FOM) and the Nederlandse Organisatie voor Wetenschappelijk Onderzoek (NWO), Netherlands;
Research Council of Norway (NFR);
Polish Ministry of Science and Higher Education;
National Authority for Scientific Research - NASR (Autoritatea Na\c{t}ional\u{a} pentru Cercetare \c{S}tiin\c{t}ific\u{a} - ANCS);
Ministry of Education and Science of Russian Federation, Russian
Academy of Sciences, Russian Federal Agency of Atomic Energy,
Russian Federal Agency for Science and Innovations and The Russian
Foundation for Basic Research;
Ministry of Education of Slovakia;
Department of Science and Technology, South Africa;
CIEMAT, EELA, Ministerio de Econom\'{i}a y Competitividad (MINECO) of Spain, Xunta de Galicia (Conseller\'{\i}a de Educaci\'{o}n),
CEA\-DEN, Cubaenerg\'{\i}a, Cuba, and IAEA (International Atomic Energy Agency);
Swedish Research Council (VR) and Knut $\&$ Alice Wallenberg
Foundation (KAW);
Ukraine Ministry of Education and Science;
United Kingdom Science and Technology Facilities Council (STFC);
The United States Department of Energy, the United States National
Science Foundation, the State of Texas, and the State of Ohio.

%% file: Alice_Authorlist_2013-Aug-20_new-CERNPREP.tex


\begingroup
\small
\begin{flushleft}
B.~Abelev\Irefn{org70}\And
J.~Adam\Irefn{org36}\And
D.~Adamov\'{a}\Irefn{org78}\And
A.M.~Adare\Irefn{org128}\And
M.M.~Aggarwal\Irefn{org82}\And
G.~Aglieri~Rinella\Irefn{org33}\And
M.~Agnello\Irefn{org88}\textsuperscript{,}\Irefn{org105}\And
A.G.~Agocs\Irefn{org127}\And
A.~Agostinelli\Irefn{org25}\And
Z.~Ahammed\Irefn{org123}
% ...
\renewcommand\labelenumi{\textsuperscript{\theenumi}~}

\section*{Affiliation notes}
\renewcommand\theenumi{\roman{enumi}}
\begin{Authlist}
\item \Adef{0}Deceased
\item \Adef{idp3002608}{Also at: Department of Applied Physics, Aligarh Muslim University, Aligarh, India}
\item \Adef{idp3720096}{Also at: M.V.Lomonosov Moscow State University, D.V.Skobeltsyn Institute of Nuclear Physics, Moscow, Russia}
\item \Adef{idp3965344}{Also at: University of Belgrade, Faculty of Physics and "Vin\v{c}a" Institute of Nuclear Sciences, Belgrade, Serbia}
\item \Adef{idp4242592}{Permanent address: Konkuk University, Seoul, Korea}
\item \Adef{idp4783920}{Also at: Institute of Theoretical Physics, University of Wroclaw, Wroclaw, Poland}
\end{Authlist}

\section*{Collaboration Institutes}
\renewcommand\theenumi{\arabic{enumi}~}
\begin{Authlist}

\item \Idef{org1}A. I. Alikhanyan National Science Laboratory (Yerevan Physics Institute) Foundation, Yerevan, Armenia
\item \Idef{org2}Benem\'{e}rita Universidad Aut\'{o}noma de Puebla, Puebla, Mexico
\item \Idef{org3}Bogolyubov Institute for Theoretical Physics, Kiev, Ukraine
\item \Idef{org4}Budker Institute for Nuclear Physics, Novosibirsk, Russia
\item \Idef{org5}California Polytechnic State University, San Luis Obispo, California, USA
\item \Idef{org6}Central China Normal University, Wuhan, China
\item \Idef{org7}Centre de Calcul de l
\item \Idef{org8}Centro de Aplicaciones Tecnol\'{o}gicas y Desarrollo Nuclear (CEADEN), Havana, Cuba
\item \Idef{org9}Centro de Investigaciones Energ\'{e}ticas Medioambientales y Tecnol\'{o}gicas (CIEMAT), Madrid, Spain
\item \Idef{org10}Centro de Investigaci\'{o}n y de Estudios Avanzados (CINVESTAV), Mexico City and M\'{e}rida, Mexico
\item \Idef{org11}Centro Fermi - Museo Storico della Fisica e Centro Studi e Ricerche ``Enrico Fermi'', Rome, Italy
\item \Idef{org12}Chicago State University, Chicago, USA
\item \Idef{org13}Commissariat \`{a} l'Energie Atomique, IRFU, Saclay, France
\item \Idef{org14}COMSATS Institute of Information Technology (CIIT), Islamabad, Pakistan
\item \Idef{org15}Departamento de F\'{\i}sica de Part\'{\i}culas and IGFAE, Universidad de Santiago de Compostela, Santiago de Compostela, Spain
\item \Idef{org16}Department of Physics Aligarh Muslim University, Aligarh, India
\item \Idef{org17}Department of Physics and Technology, University of Bergen, Bergen, Norway
\item \Idef{org18}Department of Physics, Ohio State University, Columbus, Ohio, USA
\item \Idef{org19}Department of Physics, Sejong University, Seoul, South Korea
\item \Idef{org20}Department of Physics, University of Oslo, Oslo, Norway
\item \Idef{org21}Dipartimento di Fisica dell
\item \Idef{org22}Dipartimento di Fisica dell'Universit\`{a} and Sezione INFN, Cagliari, Italy
\item \Idef{org23}Dipartimento di Fisica dell'Universit\`{a} and Sezione INFN, Trieste, Italy
\item \Idef{org24}Dipartimento di Fisica dell'Universit\`{a} and Sezione INFN, Turin, Italy
\item \Idef{org25}Dipartimento di Fisica e Astronomia dell'Universit\`{a} and Sezione INFN, Bologna, Italy
\item \Idef{org26}Dipartimento di Fisica e Astronomia dell'Universit\`{a} and Sezione INFN, Catania, Italy
\item \Idef{org27}Dipartimento di Fisica e Astronomia dell'Universit\`{a} and Sezione INFN, Padova, Italy
\item \Idef{org28}Dipartimento di Fisica `E.R.~Caianiello' dell'Universit\`{a} and Gruppo Collegato INFN, Salerno, Italy
\item \Idef{org29}Dipartimento di Scienze e Innovazione Tecnologica dell'Universit\`{a} del Piemonte Orientale and Gruppo Collegato INFN, Alessandria, Italy
\item \Idef{org30}Dipartimento Interateneo di Fisica `M.~Merlin' and Sezione INFN, Bari, Italy
\item \Idef{org31}Division of Experimental High Energy Physics, University of Lund, Lund, Sweden
\item \Idef{org32}Eberhard Karls Universit\"{a}t T\"{u}bingen, T\"{u}bingen, Germany
\item \Idef{org33}European Organization for Nuclear Research (CERN), Geneva, Switzerland
\item \Idef{org34}Faculty of Engineering, Bergen University College, Bergen, Norway
\item \Idef{org35}Faculty of Mathematics, Physics and Informatics, Comenius University, Bratislava, Slovakia
\item \Idef{org36}Faculty of Nuclear Sciences and Physical Engineering, Czech Technical University in Prague, Prague, Czech Republic
\item \Idef{org37}Faculty of Science, P.J.~\v{S}af\'{a}rik University, Ko\v{s}ice, Slovakia
\item \Idef{org38}Frankfurt Institute for Advanced Studies, Johann Wolfgang Goethe-Universit\"{a}t Frankfurt, Frankfurt, Germany
\item \Idef{org39}Gangneung-Wonju National University, Gangneung, South Korea
\item \Idef{org40}Helsinki Institute of Physics (HIP), Helsinki, Finland
\item \Idef{org41}Hiroshima University, Hiroshima, Japan
\item \Idef{org42}Indian Institute of Technology Bombay (IIT), Mumbai, India
\item \Idef{org43}Indian Institute of Technology Indore, India (IITI)
\item \Idef{org44}Inha University, College of Natural Sciences
\item \Idef{org45}Institut de Physique Nucleaire d'Orsay (IPNO), Universite Paris-Sud, CNRS-IN2P3, Orsay, France
\item \Idef{org46}Institut f\"{u}r Informatik, Johann Wolfgang Goethe-Universit\"{a}t Frankfurt, Frankfurt, Germany
\item \Idef{org47}Institut f\"{u}r Kernphysik, Johann Wolfgang Goethe-Universit\"{a}t Frankfurt, Frankfurt, Germany
\item \Idef{org48}Institut f\"{u}r Kernphysik, Technische Universit\"{a}t Darmstadt, Darmstadt, Germany
\item \Idef{org49}Institut f\"{u}r Kernphysik, Westf\"{a}lische Wilhelms-Universit\"{a}t M\"{u}nster, M\"{u}nster, Germany
\item \Idef{org50}Institut Pluridisciplinaire Hubert Curien (IPHC), Universit\'{e} de Strasbourg, CNRS-IN2P3, Strasbourg, France
\item \Idef{org51}Institute for High Energy Physics, Protvino, Russia
\item \Idef{org52}Institute for Nuclear Research, Academy of Sciences, Moscow, Russia
\item \Idef{org53}Institute for Subatomic Physics of Utrecht University, Utrecht, Netherlands
\item \Idef{org54}Institute for Theoretical and Experimental Physics, Moscow, Russia
\item \Idef{org55}Institute of Experimental Physics, Slovak Academy of Sciences, Ko\v{s}ice, Slovakia
\item \Idef{org56}Institute of Physics, Academy of Sciences of the Czech Republic, Prague, Czech Republic
\item \Idef{org57}Institute of Physics, Bhubaneswar, India
\item \Idef{org58}Institute of Space Science (ISS), Bucharest, Romania
\item \Idef{org59}Instituto de Ciencias Nucleares, Universidad Nacional Aut\'{o}noma de M\'{e}xico, Mexico City, Mexico
\item \Idef{org60}Instituto de F\'{\i}sica, Universidad Nacional Aut\'{o}noma de M\'{e}xico, Mexico City, Mexico
\item \Idef{org61}iThemba LABS, National Research Foundation, Somerset West, South Africa
\item \Idef{org62}Joint Institute for Nuclear Research (JINR), Dubna, Russia
\item \Idef{org63}Korea Institute of Science and Technology Information, Daejeon, South Korea
\item \Idef{org64}KTO Karatay University, Konya, Turkey
\item \Idef{org65}Laboratoire de Physique Corpusculaire (LPC), Clermont Universit\'{e}, Universit\'{e} Blaise Pascal, CNRS--IN2P3, Clermont-Ferrand, France
\item \Idef{org66}Laboratoire de Physique Subatomique et de Cosmologie (LPSC), Universit\'{e} Joseph Fourier, CNRS-IN2P3, Institut Polytechnique de Grenoble, Grenoble, France
\item \Idef{org67}Laboratori Nazionali di Frascati, INFN, Frascati, Italy
\item \Idef{org68}Laboratori Nazionali di Legnaro, INFN, Legnaro, Italy
\item \Idef{org69}Lawrence Berkeley National Laboratory, Berkeley, California, USA
\item \Idef{org70}Lawrence Livermore National Laboratory, Livermore, California, USA
\item \Idef{org71}Moscow Engineering Physics Institute, Moscow, Russia
\item \Idef{org72}National Centre for Nuclear Studies, Warsaw, Poland
\item \Idef{org73}National Institute for Physics and Nuclear Engineering, Bucharest, Romania
\item \Idef{org74}National Institute of Science Education and Research, Bhubaneswar, India
\item \Idef{org75}Niels Bohr Institute, University of Copenhagen, Copenhagen, Denmark
\item \Idef{org76}Nikhef, National Institute for Subatomic Physics, Amsterdam, Netherlands
\item \Idef{org77}Nuclear Physics Group, STFC Daresbury Laboratory, Daresbury, United Kingdom
\item \Idef{org78}Nuclear Physics Institute, Academy of Sciences of the Czech Republic, \v{R}e\v{z} u Prahy, Czech Republic
\item \Idef{org79}Oak Ridge National Laboratory, Oak Ridge, Tennessee, USA
\item \Idef{org80}Petersburg Nuclear Physics Institute, Gatchina, Russia
\item \Idef{org81}Physics Department, Creighton University, Omaha, Nebraska, USA
\item \Idef{org82}Physics Department, Panjab University, Chandigarh, India
\item \Idef{org83}Physics Department, University of Athens, Athens, Greece
\item \Idef{org84}Physics Department, University of Cape Town, Cape Town, South Africa
\item \Idef{org85}Physics Department, University of Jammu, Jammu, India
\item \Idef{org86}Physics Department, University of Rajasthan, Jaipur, India
\item \Idef{org87}Physikalisches Institut, Ruprecht-Karls-Universit\"{a}t Heidelberg, Heidelberg, Germany
\item \Idef{org88}Politecnico di Torino, Turin, Italy
\item \Idef{org89}Purdue University, West Lafayette, Indiana, USA
\item \Idef{org90}Pusan National University, Pusan, South Korea
\item \Idef{org91}Research Division and ExtreMe Matter Institute EMMI, GSI Helmholtzzentrum f\"ur Schwerionenforschung, Darmstadt, Germany
\item \Idef{org92}Rudjer Bo\v{s}kovi\'{c} Institute, Zagreb, Croatia
\item \Idef{org93}Russian Federal Nuclear Center (VNIIEF), Sarov, Russia
\item \Idef{org94}Russian Research Centre Kurchatov Institute, Moscow, Russia
\item \Idef{org95}Saha Institute of Nuclear Physics, Kolkata, India
\item \Idef{org96}School of Physics and Astronomy, University of Birmingham, Birmingham, United Kingdom
\item \Idef{org97}Secci\'{o}n F\'{\i}sica, Departamento de Ciencias, Pontificia Universidad Cat\'{o}lica del Per\'{u}, Lima, Peru
\item \Idef{org98}Sezione INFN, Bari, Italy
\item \Idef{org99}Sezione INFN, Bologna, Italy
\item \Idef{org100}Sezione INFN, Cagliari, Italy
\item \Idef{org101}Sezione INFN, Catania, Italy
\item \Idef{org102}Sezione INFN, Padova, Italy
\item \Idef{org103}Sezione INFN, Rome, Italy
\item \Idef{org104}Sezione INFN, Trieste, Italy
\item \Idef{org105}Sezione INFN, Turin, Italy
\item \Idef{org106}SUBATECH, Ecole des Mines de Nantes, Universit\'{e} de Nantes, CNRS-IN2P3, Nantes, France
\item \Idef{org107}Suranaree University of Technology, Nakhon Ratchasima, Thailand
\item \Idef{org108}Technical University of Split FESB, Split, Croatia
\item \Idef{org109}The Henryk Niewodniczanski Institute of Nuclear Physics, Polish Academy of Sciences, Cracow, Poland
\item \Idef{org110}The University of Texas at Austin, Physics Department, Austin, TX, USA
\item \Idef{org111}Universidad Aut\'{o}noma de Sinaloa, Culiac\'{a}n, Mexico
\item \Idef{org112}Universidade de S\~{a}o Paulo (USP), S\~{a}o Paulo, Brazil
\item \Idef{org113}Universidade Estadual de Campinas (UNICAMP), Campinas, Brazil
\item \Idef{org114}University of Houston, Houston, Texas, USA
\item \Idef{org115}University of Jyv\"{a}skyl\"{a}, Jyv\"{a}skyl\"{a}, Finland
\item \Idef{org116}University of Liverpool, Liverpool, United Kingdom
\item \Idef{org117}University of Tennessee, Knoxville, Tennessee, USA
\item \Idef{org118}University of Tokyo, Tokyo, Japan
\item \Idef{org119}University of Tsukuba, Tsukuba, Japan
\item \Idef{org120}University of Zagreb, Zagreb, Croatia
\item \Idef{org121}Universit\'{e} de Lyon, Universit\'{e} Lyon 1, CNRS/IN2P3, IPN-Lyon, Villeurbanne, France
\item \Idef{org122}V.~Fock Institute for Physics, St. Petersburg State University, St. Petersburg, Russia
\item \Idef{org123}Variable Energy Cyclotron Centre, Kolkata, India
\item \Idef{org124}Vestfold University College, Tonsberg, Norway
\item \Idef{org125}Warsaw University of Technology, Warsaw, Poland
\item \Idef{org126}Wayne State University, Detroit, Michigan, USA
\item \Idef{org127}Wigner Research Centre for Physics, Hungarian Academy of Sciences, Budapest, Hungary
\item \Idef{org128}Yale University, New Haven, Connecticut, USA
\item \Idef{org129}Yonsei University, Seoul, South Korea
\item \Idef{org130}Zentrum f\"{u}r Technologietransfer und Telekommunikation (ZTT), Fachhochschule Worms, Worms, Germany
\end{Authlist}
\endgroup